# ACMI: An index for exposed coal mapping using Landsat imagery

Zhen Yang[a], Jianyong Zhang[b,*], Yanchuang Zhao[a]

[a] *College of Information Science and Engineering, Henan University of Technology, Zhengzhou 450001, China*;

[b] *College of Earth Science, Chengdu University of Technology, Chengdu 610059, China*;

**Abstract:** Remotely sensing the spatial distribution of exposed coal (EC) is significant for understanding the footprints of mining activities. However, widely applicable methods for the identification of EC surfaces remain inadequate because the choices of recent methods confront the diverse EC types and backgrounds. Therefore, this study proposed a new Automated Coal Mapping Index (ACMI) which was empirically formulated by an iterative process of identifying parameters that maximize the separability of EC and non-EC surfaces. The performance of ACMI was tested in six study areas worldwide with different landscape types and coal types. Based on the visual inspection, ACMI was more effective in highlighting EC surfaces and suppressing non-EC surfaces than the existing methods. Compared with the sample points obtained through direct interpretation, ACMI obtained better EC mapping results than previous methods with the F1 score and overall accuracy (OA) no less than 0.91 and 93.20% across all the selected Landsat images of the study areas, respectively. In addition, ACMI was demonstrated to have a stable optimal threshold and 0 can serve as its default threshold. The default threshold makes EC mapping using ACMI an automated process. The new index has the potential to support a variety of mining-activity-related studies, such as the identification of mining disturbances and illegal mining detection at multi-spatial-temporal scales.

## 1. Introduction

Coal is one of the most widely used fossil energy sources in the world [1], which has contributed greatly to the development of human society [2], [3] and has played an important role in fields such as electricity, industrial production, heating, and chemical engineering. In the past few decades, coal has always accounted for a large proportion (>20%) of the global energy structure [4]. In 2022, global coal consumption exceeded 8 billion tons for the first time [5], and coal production showed a significant increase in some major coal-producing countries such as China, India, and Indonesia.

Although coal has provided great energy support for human society, land cover changes caused by coal mining negatively influence the ecological environment, agricultural production, and living conditions of residents [6], [7]. In some areas, coal mining may even lead to social problems, as decisions regarding the benefits of protecting the land and those of growing the economy may be in conflict as a result of different viewpoints held by miners, local residents, and government officials [6]. Thus, much public attention has been focused on mining issues and the need to detect the spatiotemporal distribution of mining activities [8], [9]. Exposed coal (EC) is the essential product of coal mining and is widely distributed in coal-mining-related regions such as coal yards in mining areas and operation areas of surface coal mines, marking it can serve as important and reliable evidence for detecting the footprints of mining activities [9], [10]. Therefore, precise identification of EC surfaces is helpful for detecting mining activities and understanding the influence of coal mining.

Satellite sensors can record the earth at various spatial and temporal resolutions, which have been extensively applied in extracting and analyzing information regarding land cover changes [11], [12]. Due to the free and open Landsat data policy, Landsat sensors have been widely applied in mining regions and other anthropic zones [13], [14], [15]. The current EC mapping methods based on Landsat images can be divided into classifier-based methods and index-based methods. The classifier-based methods adopt sample training to generate prior knowledge for supervised algorithms, such as the random forest classifier [9], machine learning algorithms [16], and subspace classifier [17] to separate EC pixels from backgrounds. The classifier-based methods can obtain desirable results in local regions, but they are not suitable for large-scale mapping because of the need for enough samples and computationally intensive models [18].

Index-based methods, which are designed to highlight the target object by combining the reflectance of two or more bands, have been conventional means of land-use change detection. In comparison with classifier-based methods, index-based methods gain significant superiorities in large-scale mapping owing to their high efficiency, low computation complexity, and convenience in practical applications [19]. Mao et al. [20] developed the normalized difference coal index (NDCI) using the reflectance difference between the near-infrared (NIR) band and shortwave infrared 1 (SWIR1) band of Landsat TM images. Mukherjee et al. [21] proposed the coal mine index (CMI) using SWIR 1 and shortwave infrared 2 (SWIR2) bands of Landsat OLI images. The CMI has achieved high accuracy for identifying EC surfaces in Jharkhand, India. Yang et al. [8] argued that EC has the unique reflectance trend reflected in NIR, SWIR1, and SWIR2 bands, and they determined the bare coal index (BCI) to successfully identify the EC distributions in a surface mining region of Inner Mongolia, China. Li et al. [10] and He et al. [22] have proved the reliability of BCI when identifying the EC distributions in different regions of the world. Pan et al. [23] developed the Exposed Coal Index (ECI) by considering the spectral characteristics reflected in shape and range. The ECI has been demonstrated to be applicable to both Sentinel-2 and Landsat OLI images.

Even though the index-based methods have achieved satisfactory results in EC mapping, the selection between them confronts the following two issues: the issue of diverse backgrounds and the issue of diverse EC spectra. As shown in Fig. 1(a), EC surfaces in different regions are always under diverse backgrounds. For instance, the EC surfaces in Xinjiang are usually with a relatively simple background, while the EC surfaces in Inner Mongolia, North Rhine-Westphalia (NRW) and Shanxi are under complex urban or rural scenes. The diversity of backgrounds requires a new EC index can be effective in suppressing the diverse background objects. In addition, as shown in Fig. 1(b), EC surfaces in different regions may vary a lot in spectral characteristics. In Inner Mongolia and Wyoming, the spectral curve of EC pixels shows a more obvious upward trend than that of Xinjiang, especially in the last three bands (NIR, SWIR1 and SWIR2 bands). In the region of NRW, the spectral curve is characterized by a downward trend in a relatively higher range at the last two bands (SWIR1 and SWIR2 bands). This makes it a challenge to present a generic index that has the characteristic of being useful for different regions.

Therefore, the aim of this study is to develop a widely applicable index based on Landsat imagery, which can achieve good EC identification results in different areas around the world. To achieve this goal, a multiple-band index named Automated Coal Mapping Index (ACMI) is proposed by analyzing the spectral signature of EC surfaces on Landsat imagery. Six study areas with different landscape types and coal types were selected to test the performance of the ACMI and evaluate its relative accuracies in comparison with other methods. Based on the qualitative and quantitative evaluation of EC mapping results, the ACMI is expected to effectively separate the EC surfaces and background land covers in different areas.

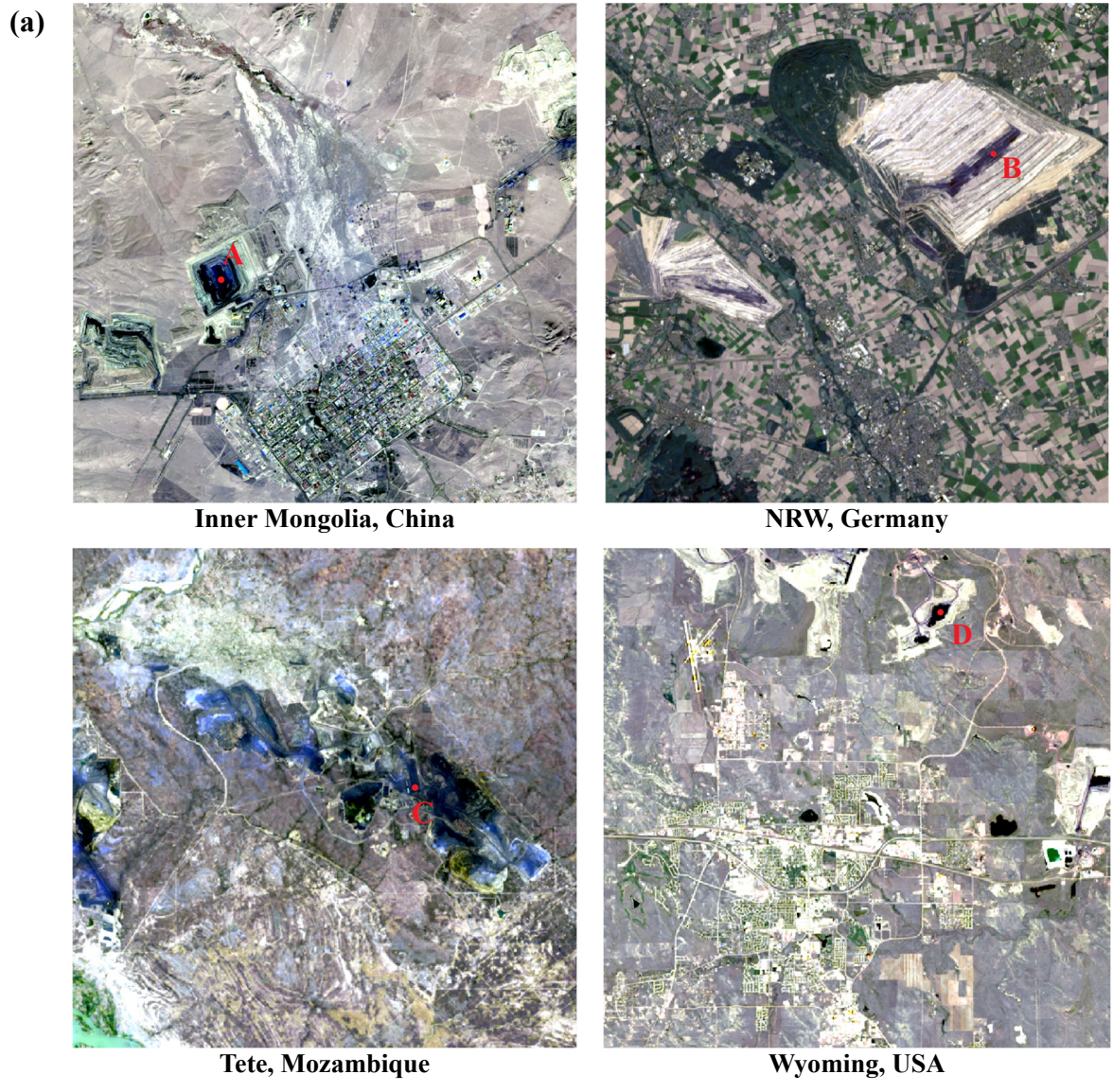

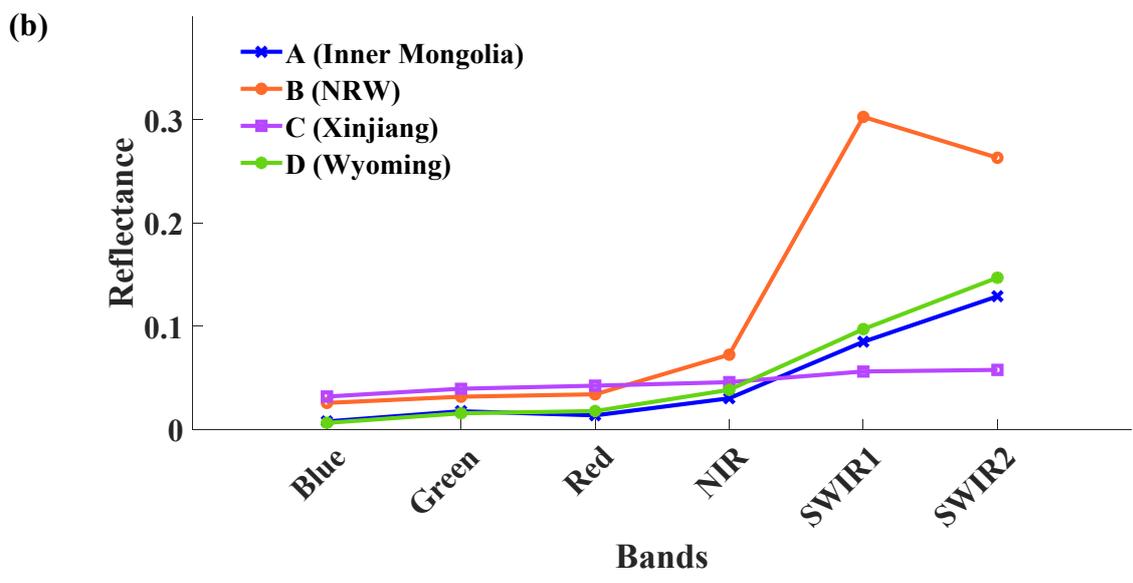

**Fig. 1.** The two major issues for EC mapping through index-based methods: (a) the issue of diverse backgrounds, and (b) the issue of diverse EC spectra.

## 2. Study areas and data

### 2.1. Study areas

We tested the ACMI in six areas (Fig. 2 and Table 1), ranging from the mid-temperate region (6) to temperate (1, 2, 4, and 5) and tropical (3) regions. All the study areas have at least one large surface coal mine and diversified land cover types. The landscape type of the study area 2 is forest, while study areas 1, 3, 4, 5, and 6 are mainly covered by shrubs, grasslands, or deserts. The selected study areas capture a diverse mix of coal types, including anthracite (2 and 4), bituminous coal (2, 3, and 5), subbituminous coal (1), and lignite (6).

The study areas were carefully selected to ensure that the sub-scenes cover various background land covers, such as dark soil, built-up areas, water bodies and other dark surfaces as background to the EC surfaces. The study areas in Wyoming, North Rhine-Westphalia (NRW), Tete and Inner Mongolia are characterized by the presence of dark soil, built-up surfaces, and water bodies. The area in Shanxi also consists of built-up surfaces and dark soil but with few water bodies. The study area in Xinjiang is characterized by a relatively simple background with massive dark soil but few water bodies and built-up surfaces.

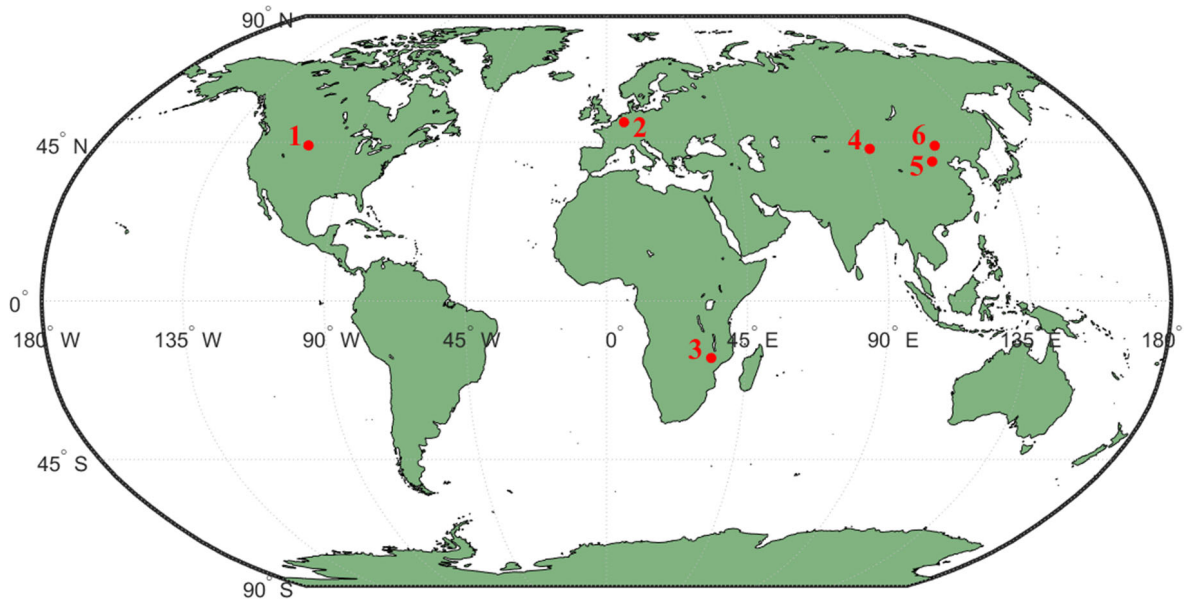

**Fig. 2.** Locations of the study areas: 1. Wyoming (USA), 2. North Rhine-Westphalia (Germany), 3. Tete (Mozambique), 4. Xinjiang (China), 5. Shanxi (China), 6. Inner Mongolia (China).

Table 1. Characteristics of the selected study areas: location, landscape type, climate zone and coal type.

| No. | Location | Landscape type | Climate zone | Coal type |
|---|---|---|---|---|
| 1 | Wyoming (USA) | Grassland | Temperate continental semi-arid climate | Subbituminous coal |
| 2 | NRW (Germany) | Forest | Temperate maritime climate | Anthracite and bituminous coal |
| 3 | Tete (Mozambique) | Grassland | Tropical savanna climate | Bituminous coal |
| 4 | Xinjiang (China) | Desert | Temperate continental arid climate | Anthracite |
| 5 | Shanxi (China) | Shrub | Temperate continental monsoon climate | Bituminous coal |
| 6 | Inner Mongolia (China) | Grassland | Mid-temperate continental semiarid climate | Lignite |

## 2.2. Landsat images

It has been reported that the surface reflectance differences between OLI and TM/ETM+ images are about 2% [24]. Therefore, an index may result in different effects between Landsat 8 and prior Landsat sensors. To test the applicability of ACMI in Landsat sensors, we have selected two Landsat surface reflectance images of L2SP (Level 2 Science Product) for each study area, one of which is from the TM/ETM+ sensors, and the other is from the Operational Land Imager (OLI) sensor. The images were acquired in a geometrically terrain-corrected state from the US Geological Survey (USGS) and were surface reflectance-corrected with Landsat surface reflectance algorithms [25], [26]. The selected images have no clouds or shadows over the study areas and their details including acquisition date, path/row in the Worldwide Reference System (WRS) and sensor type are shown in Table 2. In addition to the Landsat images of the six study areas by which accuracy evaluations and comparisons were fulfilled, validation of the anti-noise ability of the new method was conducted using a cloud-contaminated Landsat image in Inner Mongolia, China. Detailed information on this additional Landsat image and its classification output is included in the Discussion section.

Table 2. Details of the selected Landsat images for the six study areas.

| No. of the study area | Path/row in WRS | Sensor type | Acquisition date (Year/month/day) |
|---|---|---|---|
| 1 | 34/29 | TM | 2008/09/19 |
|   |       | OLI | 2021/06/03 |
| 2 | 197/25 | TM | 2008/05/04 |
|   |        | OLI | 2019/08/23 |
| 3 | 168/71 | ETM+ | 2021/11/29 |
|   |        | OLI | 2020/09/15 |
| 4 | 139/30 | ETM+ | 2016/07/15 |
|   |        | OLI | 2020/09/20 |
| 5 | 126/33 | TM | 2011/04/10 |
|   |        | OLI | 2016/07/28 |
| 6 | 124/29 | TM | 2010/08/30 |
|   |        | OLI | 2022/10/19 |

## 3. Method

### 3.1. Spectral analysis of land covers

A reliable EC index should have the ability to accurately discriminate between non-EC surfaces and EC surfaces. To achieve this goal, the reflectance values of different land cover types were analyzed based on the selected Landsat TM/ETM+ images (Table 2). To include all coal types, 25 EC pixels were selected from each TM/ETM+ image, and this eventually produced 150 EC pixels. The reflectance values of non-EC surfaces were sampled from the Landsat TM image of the study area in Inner Mongolia. The major land cover types of non-EC in this image are water, vegetation, bright soil, brown soil, dark soil, bright built-up surfaces (BUS), dark BUS, and red BUS. The reason for selecting this image for spectral analysis of non-EC was that this image consists of all the major challenging elements affecting EC mapping accuracy: dark BUS, dark soil, and other low-reflectivity surfaces such as water bodies. For each of the above land cover types, 150 pixels were extracted and the reflectance distributions of each band for major land cover types including EC are shown in Fig. 3. Using the Jeffries–Matusita's interclass separability measure [27], all pairs of the nine major land cover types were demonstrated to be separable with pair separation values ranging from 1.84 to 2.00.

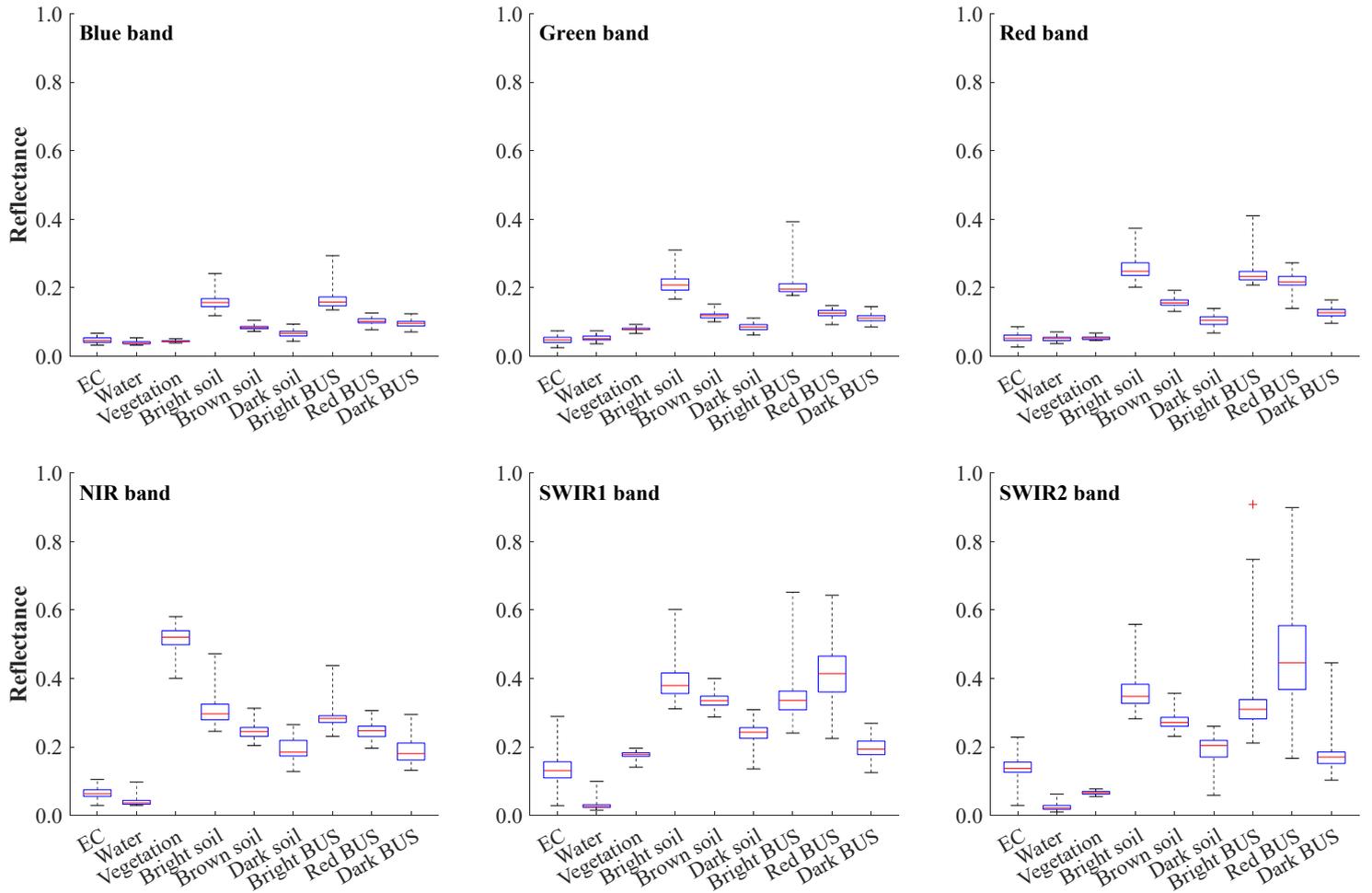

**Fig. 3.** The distributions of pixel reflectance for different land cover types. Box plots represent the 25th, 50th, and 75th percentiles, and the whiskers extend to the maximum and minimum.

### 3.2. Formulation of the Automated Coal Mapping Index (ACMI)

Based on the formulation procedure of the automated water extraction index (AWEI) described in Feyisa et al. [28], a new index (ACMI) was developed using six spectral bands of Landsat images to enlarge the difference between EC and other land covers. Accordingly, an equation is formulated to effectively suppress non-EC surfaces and extract EC pixels with high accuracy by band differencing and coefficient setting (Eq. (1)). Using the samples of different land cover types selected in Section 3.1, the coefficients of (1) and the selection of Landsat bands are empirically determined based on an iterative process of identifying parameters that maximize the separability of EC and non-EC surfaces. In this iterative process, special attention was paid to enhancing the separability of dark backgrounds such as dark soil and dark building that are hard to discriminate because of similarities in reflectance signatures. Besides, the determination of the coefficients was also to make sure a stable threshold of EC extraction by forcing non-EC pixels below 0 and EC pixels above 0, indicating that 0 can serve as the default threshold for classifying Landsat pixels into EC and non-EC under different coal types and various background environments.

$$\text{ACMI} = \begin{cases} 4.75 \times \rho_{blue} - \rho_{green} - 4.5 \times \rho_{NIR} + 0.25 \times \rho_{SWIR1} + \rho_{SWIR2} + 0.1 \\ -1, \quad \text{MNDWI} > 0 \\ -1, \quad Max(\rho_{blue}, \rho_{green}, \rho_{red}) > 0.075 \end{cases} \quad (1)$$

where $\rho$ is the reflectance value of blue, green, NIR, SWIR1 and SWIR2 bands of Landsat images; *Max* is the maximum function; MNDWI represents the normalized difference water index developed by developed by Xu [29] as

$$\text{MNDWI} = \frac{\rho_{green} - \rho_{SWIR1}}{\rho_{green} + \rho_{SWIR1}} \quad (2)$$

In Eq. (1), using $4.75 \times \rho_{blue}$ subtract $4.5 \times \rho_{NIR}$ and $\rho_{green}$ leads to relatively large negative values for most bright surfaces and values closer to 0 for most dark pixels. To assist in distinguishing EC from other land covers of similar spectral patterns, SWIR1 and SWIR2 bands are adopted to balance the result and different coefficients are set for these two bands to make sure EC pixels have positive values and most non-EC surfaces are negative. On the other hand, water bodies in the result are further suppressed using MNDWI with 0 as the threshold, and bright surfaces are further suppressed using 0.1 as the maximum value of the reflectance of visible bands (red, green, and blue bands).

Because of the phenomenon of different objects with the same spectrum, the identification results of target objects generated from remote sensing images may be mixed with some pixels of other land cover types [30]. Our study was confronted with this problem as some pixels in towns or on the shoreside of rivers have almost the same spectral characteristics as EC. These pixels are sporadic and discontinuous, and therefore median filtering with a sliding window size of 3×3 is used to deal with the EC extraction results of ACMI to suppress these pixels.

### 3.4. Validation

*3.4.1. Comparison with existing methods*

To compare the performance of the ACMI with existing methods, we conducted preliminary evaluations of EC indices including the NDCI [20], BCI [8], CMI [21], and ECI [23]. Based on the preliminary evaluations, NDCI and CMI failed to effectively suppress multiple land cover types such as water, soil, and built-up surfaces in our study areas. BCI and ECI showed much better performance in suppressing non-EC surfaces and highlighting EC surfaces. Compared with ECI, BCI and the new index have the same advantage of not requiring sample training to determine the threshold of EC mapping. Thus, we only considered BCI for comparison with the new proposed index. BCI classifies pixels into binary classes with 1 for EC and 0 for non-EC, which can be expressed as

$$\mathrm{BCI} = \begin{cases} 1, & \text{if } \rho_{\mathrm{NIR}} < \rho_{\mathrm{SWIR1}} < \rho_{\mathrm{SWIR2}} < 0.15 \\ 0, & \text{else} \end{cases} \quad (3)$$

For the sake of fairness in comparison, the EC extraction results of BCI were also filtered using the same median filtering as that adopted in the EC extraction results of ACMI. For ACMI, the EC mapping results used for evaluation and comparison were generated using 0 as the threshold.

### 3.4.2. Qualitative evaluation

Qualitative evaluation intends for a direct assessment of the ability of the two methods to highlight EC surfaces and suppress background land covers. To make it easy to capture EC distributions from the results, the pixels identified as EC are presented in black while the background pixels are displayed in light gray. Subsequently, all results are evaluated by visually inspecting the original Landsat images. In addition, the results are evaluated in a large area and a local region with some commission errors and omission errors marked, so as to clearly evaluate and compare the overall and detailed performance of the two methods.

### 3.4.3. Quantitative evaluation

Ideally, maps generated from remote sensing images should be evaluated by independent validation data. However, no independent samples of EC surfaces and other land covers are available for the study areas. The most credible source for this accuracy assessment was the remote sensing images themselves when no independent validation data were available [31]. In addition, the pixels of background land covers dominate the study areas, which will affect the objectivity of accuracy assessment if randomly selecting samples in the whole areas [32], [33]. Therefore, it is necessary to ensure that the target object, i.e., the EC surface, is fully reflected in the samples.

In this study, the classification accuracies of the two methods were determined using a quantitative procedure based on randomly selected samples. Firstly, the EC polygons within a test image were captured from the image by direct interpretation with the help of high-resolution Google Earth images. For each image, 300 samples were randomly selected within the EC polygons, and 450 background samples were randomly selected outside the EC polygons. This produces 9000 samples with 3600 samples for EC and the rest for background land covers. Finally, based on the selected samples, EC mapping results were evaluated using overall accuracy (OA), producer's accuracy (PA), user's accuracy (UA), and F1 score. The F1 score can be computed as

$$F1 = 2 \times \frac{UA \times PA}{UA + PA} \quad (4)$$

## 4. Results

### 4.1. Qualitative evaluation results

*4.1.1. EC mapping results of Wyoming*

The EC mapping results of the two methods in Wyoming are shown in Fig. 4. On the whole, as shown in Fig. 4(b), (c), (h) and (i), the two methods can effectively suppress the backgrounds and the EC surfaces extracted by ACMI are much larger than those obtained by BCI for both the TM and OLI images. In the detailed results of the TM image (Fig. 4(d), (e) and (f)), the water body in the blue circle was perfectly suppressed by the two methods. In Fig. 4(e), many EC pixels of the TM image were wrongly classified as non-EC by BCI. ACMI appeared to be more effective in identifying the EC surfaces of the TM image and got a more accurate EC mapping result than BCI in terms of shape and continuity (Fig. 4(f)). By comparing Fig. 4(k) and (l), ACMI was found to be more effective in identifying EC surfaces than BCI. It should be noted that most linear EC surfaces were misclassified by the two methods, and we will try to analyze the reasons for such misclassifications in the Discussion.

*4.1.2. EC mapping results of NRW*

The EC mapping results of the two methods in NRW are shown in Fig. 5. As shown in Fig. 5(b), (e), (h) and (k), BCI failed to extract almost all EC pixels in both the TM and OLI images, even though nearly all backgrounds were suppressed. This is because almost all EC pixels in this area show a common reflectance trend of $\rho_{SWIR1} > \rho_{SWIR2}$ that makes them fail to match the rule of BCI (Eq. (3)). In contrast, ACMI showed a clear advantage in identifying EC pixels while suppressing background land covers for both the TM and OLI images (Fig. 5(c), (f), (i) and (l)).

*4.1.3. EC mapping results of Tete*

The EC mapping results of the two methods in Tete are shown in Fig. 6. As shown in Fig. 6(b) and (c), EC surfaces outputted by ACMI for the ETM+ image are of a larger extent than that extracted by BCI. It is mainly because BCI seemed to omit some EC pixels and ACMI obtained a much more precise EC mapping result (Fig. 6(e) and (f)). Nevertheless, the two methods achieved similar overall results for the OLI image (Fig. 6(h) and (i)). In the detailed results of the OLI image (Fig. 6(j), (k) and (l)), the water body in the green ellipse was effectively suppressed by the two methods. As shown in the blue ellipse in Fig. 6 (j), (k) and (l), ACMI still showed a clear advantage in extracting EC surfaces in this area.

*4.1.4. EC mapping results of Xinjiang*

The EC mapping results of the two methods in Xinjiang are shown in Fig. 7. As shown in Fig. 7(b),

(c), (h) and (i), the two methods achieved similar overall results for both the ETM+ and OLI images. In detailed results of the ETM+ image (Fig. 7(e) and (f)), ACMI and BCI showed good performances in suppressing various background land covers including dark soil, built-up surfaces, and the water body in the green circle of Fig. 7 (d). As clearly shown in Fig. 7(j), (k) and (l), the two methods also effectively suppressed background land covers and ACMI showed a clear advantage in identifying EC surfaces. However, the EC surface in the red circle of Fig. 7(j) was misclassified by the two methods.

*4.1.5. EC mapping results of Shanxi*

The EC mapping results of the two methods in Shanxi are shown in Fig. 8. As shown in Fig 8(b), (c), (h) and (i), for both the TM and OLI images, EC surfaces extracted by BCI are slightly smaller than those obtained by ACMI. This is because some EC pixels were misclassified as non-EC by BCI (Fig. 8(e) and (k)), such as some EC surfaces in the red ellipse of Fig. 8(d) and the blue circle of Fig. 8(j). In contrast, as shown in Fig. 8(f) and (l), ACMI performed well in extracting EC surfaces. In addition, the two methods can effectively suppress background land covers such as the dark soil shown in the green ellipse of Fig. 8(d).

*4.1.6. EC mapping results of Inner Mongolia*

The EC mapping results of the two methods in Inner Mongolia are shown in Fig. 9. On a large scale, as shown in Fig. 9(b), (c), (h) and (i), EC surfaces extracted by BCI for both the TM and OLI images are smaller than those obtained by ACMI, largely because of the omissions of some EC pixels in BCI [shown as the blue ellipses in Fig. 9(e) and (k)]. In the detail results, as shown in Fig. 9(e), (f), (k) and (l), the two methods effectively suppressed background land covers. However, as shown in the blue ellipses of Fig. 9(e) and (k), BCI failed to extract some EC pixels, pointing to the bad performance of BCI in some regions of this study area. As shown in Fig. 9(f) and (l), ACMI showed a clear advantage in identifying EC surfaces, even if some EC pixels in the OLI image were omitted by ACMI (shown as the green circle of Fig. 9(j) and (l)).

According to the above analyses on the EC mapping results of the two methods, ACMI is concluded to have the following advantages. First, ACMI showed the similar excellent performance as BCI in suppressing various background land covers in all six study areas. Second, ACMI displayed high effectiveness in EC mapping across all six study areas. The BCI, on the contrary, was only effective in some study areas but failed to be useful for all. For instance, BCI achieved good results in Xinjiang and Shanxi, but it almost misclassified all the EC pixels in NRW.

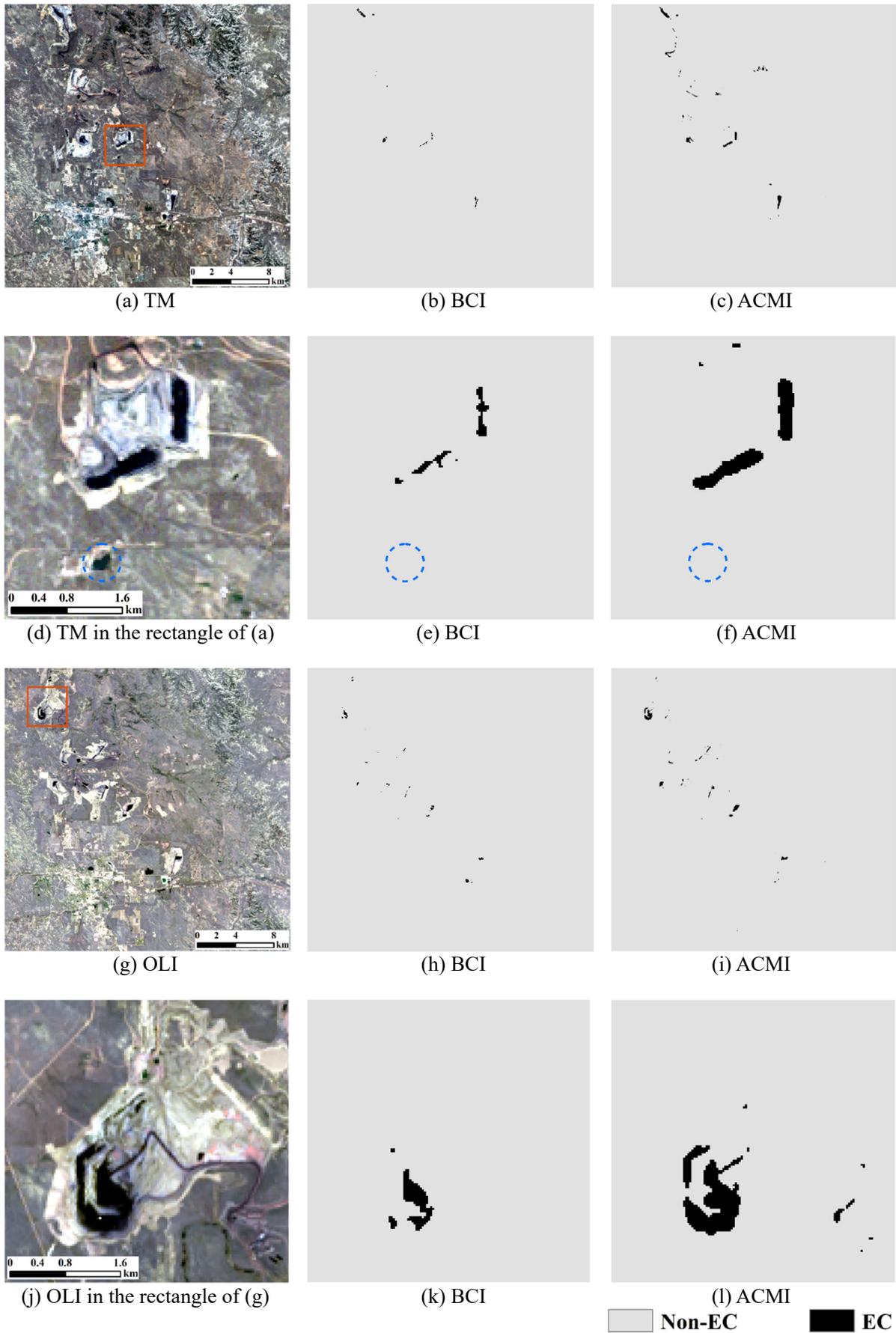

**Fig. 4.** EC mapping results of Wyoming.

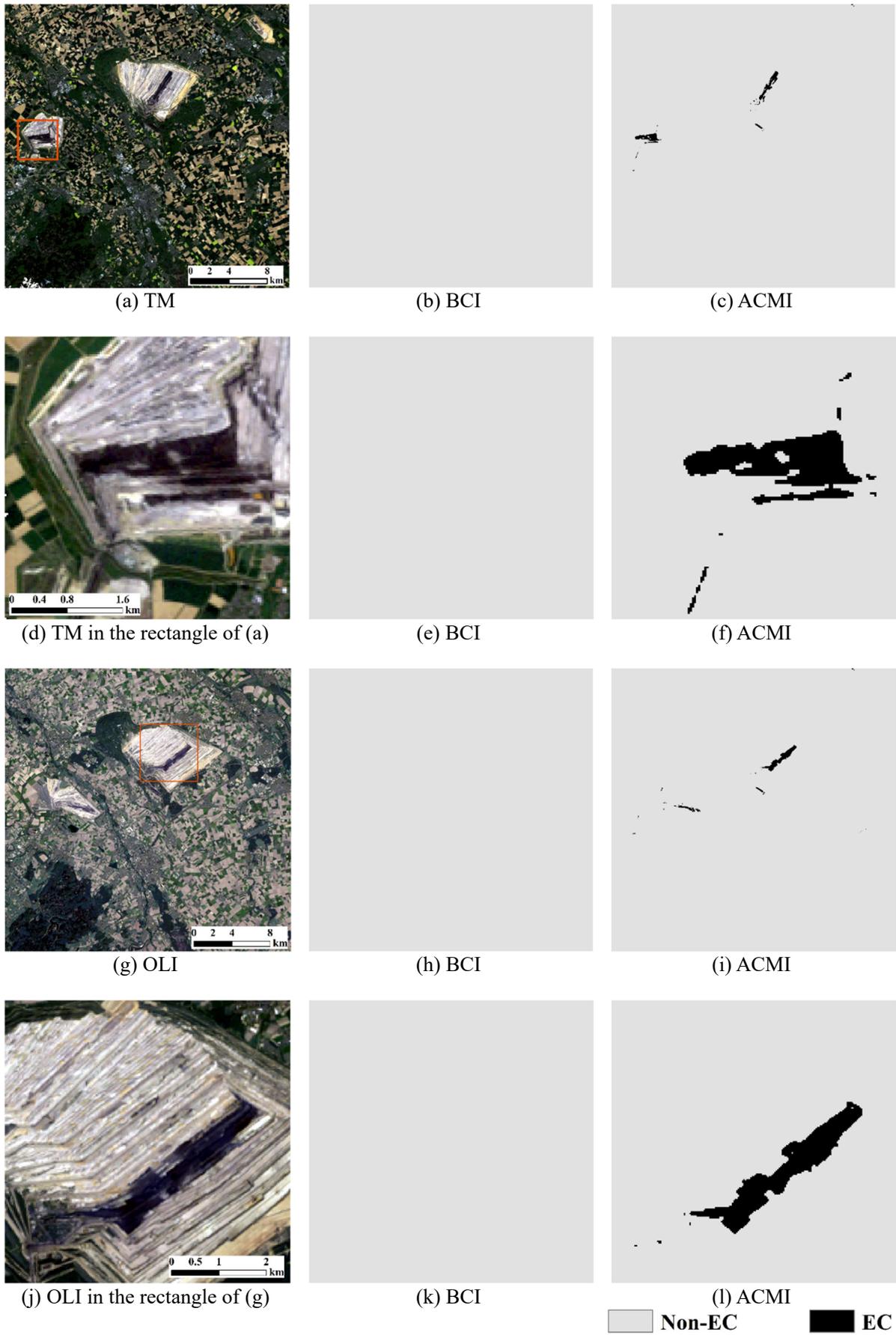

Fig. 5. EC mapping results of NRW.

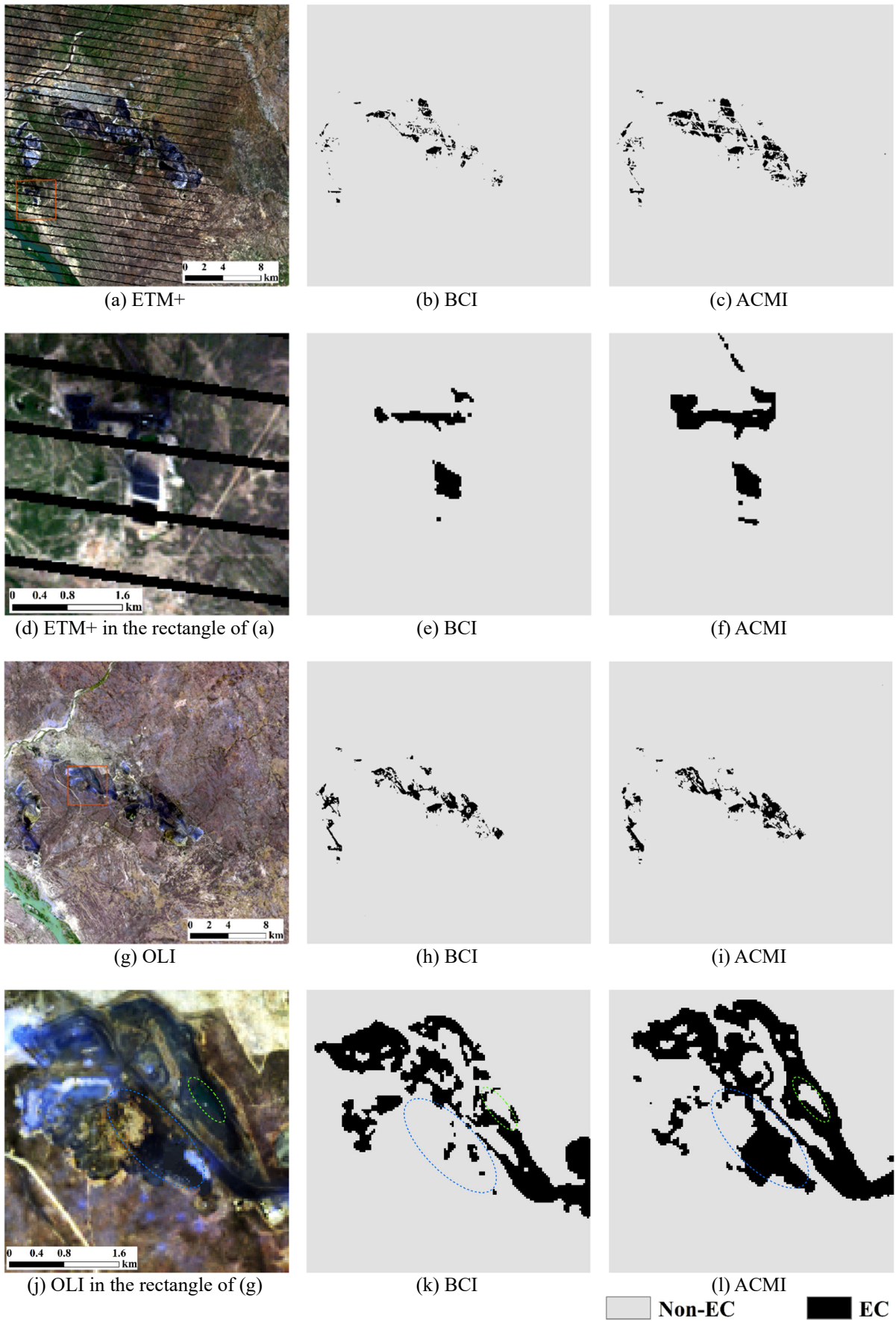

**Fig. 6.** EC mapping results of Tete.

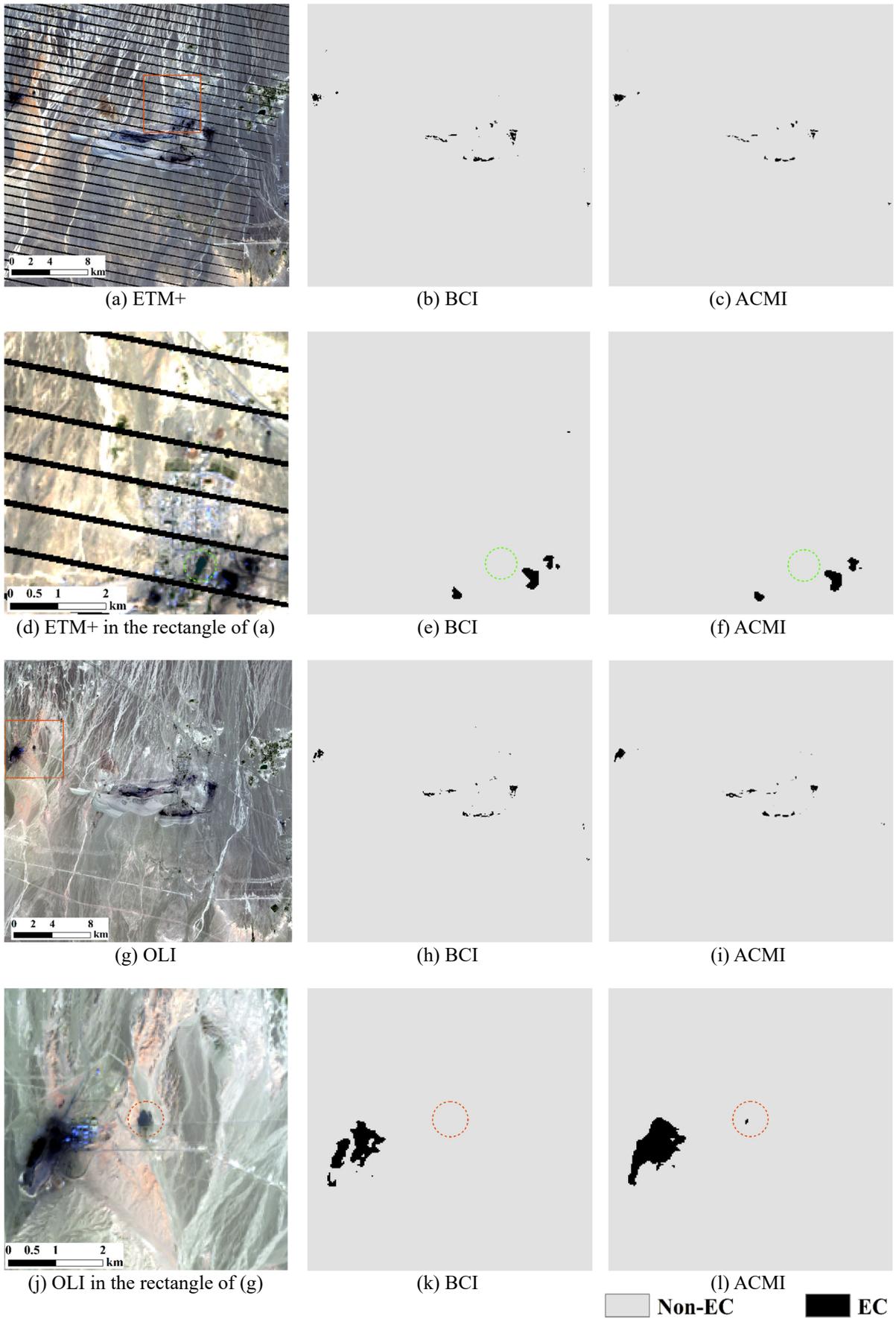

**Fig. 7.** EC mapping results of Xinjiang.

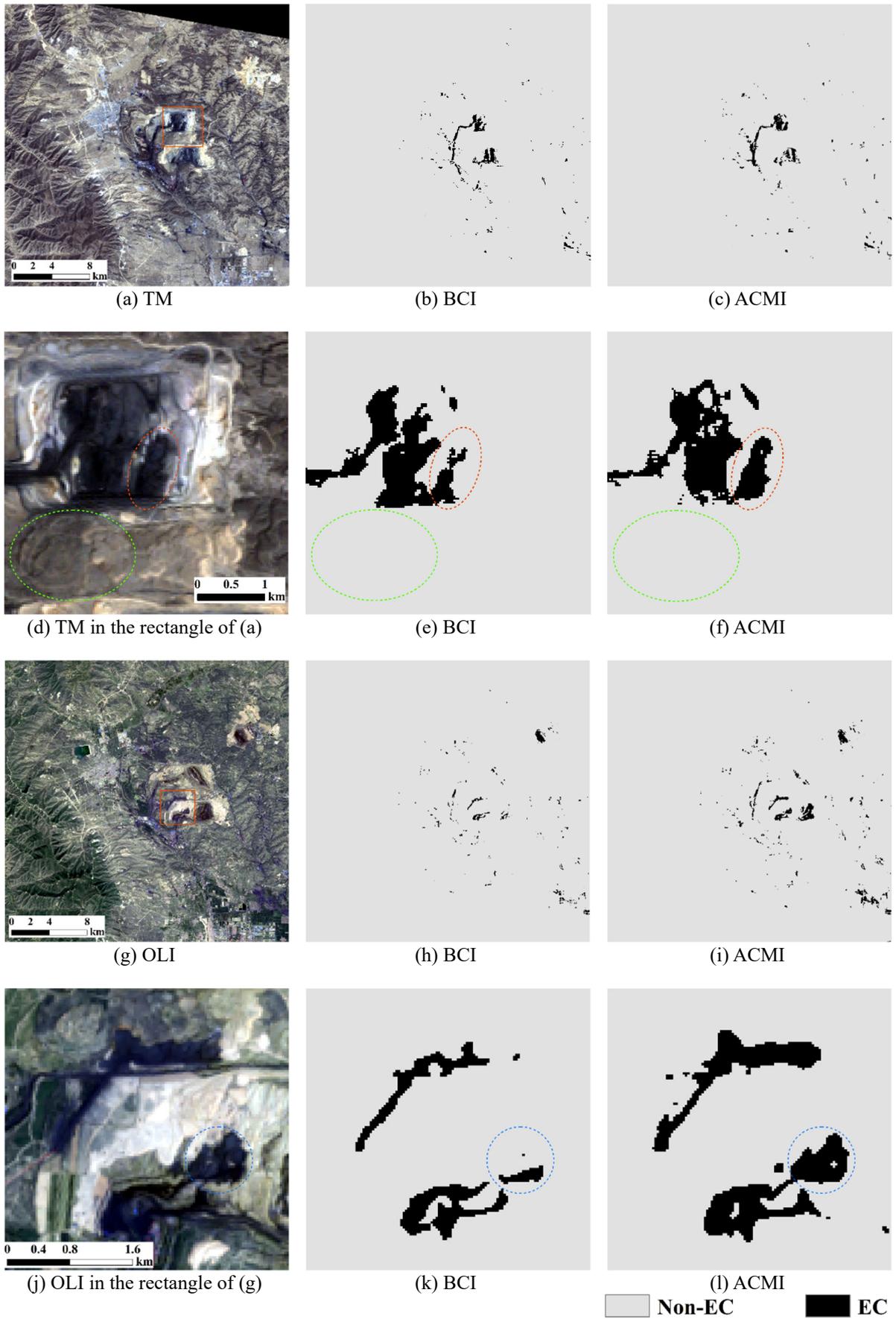

**Fig. 8.** EC mapping results of Shanxi.

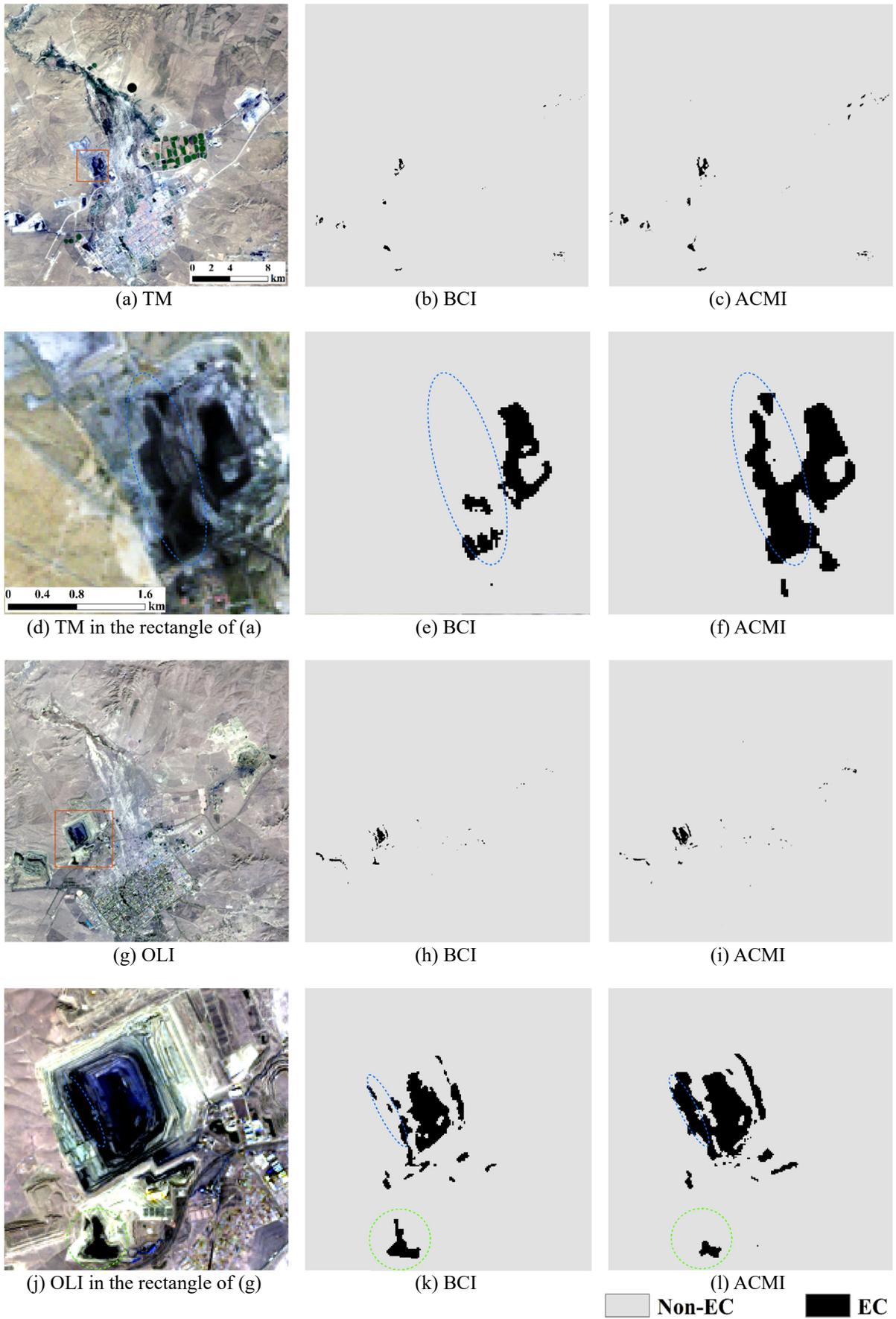

**Fig. 9.** EC mapping results of Inner Mongolia.

## 4.2. Quantitative evaluation results

The results of EC extraction accuracy of the two classification methods at each of the six study areas are summarized in Table 3. The UA, PA, F1 score and OA of ACMI ranged from 95.88% to 100.00%, 83.00% to 99.00%, 0.91 to 0.99, and 93.20% to 99.20%, respectively. At all study areas, the F1 score and OA achieved by ACMI for both TM/ETM+ and OLI images were higher than that of BCI. The average UA, PA, F1 score and OA of ACMI were 99.24%, 92.50%, 0.96 and 96.70%, respectively; all of them are higher than those achieved by BCI, indicating that ACMI performed better than BCI at extracting EC and suppressing non-EC surfaces on the whole. Especially, at the study area in NRW, the PA of BCI for both TM and OLI images are 0.00%, showing that BCI failed to identify EC surfaces from the backgrounds. In this study area, all accuracy metrics of ACMI are no less than 90.00%. However, at the study area in Xinjiang, the PA of ACMI for the ETM+ image is slightly lower than that of BCI, proving that ACMI omitted more EC pixels than BCI in this image.

Table 3. Accuracies of EC mapping results using ACMI and BCI.

| Study area | Sensor | Method | UA (%) | PA (%) | F1 (%) | OA (%) |
|---|---|---|---|---|---|---|
| Wyoming | TM | ACMI | 100.00 | 90.00 | 94.74 | 96.00 |
|  |  | BCI | 100.00 | 33.00 | 49.62 | 73.20 |
|  | OLI | ACMI | 100.00 | 95.00 | 97.44 | 98.00 |
|  |  | BCI | 97.92 | 47.00 | 63.51 | 78.40 |
| NRW | TM | ACMI | 100.00 | 90.00 | 94.74 | 96.00 |
|  |  | BCI | - | 0.00 | 0.00 | 60.00 |
|  | OLI | ACMI | 98.97 | 96.00 | 97.46 | 98.00 |
|  |  | BCI | - | 0.00 | 0.00 | 60.00 |
| Tete | ETM+ | ACMI | 98.02 | 99.00 | 98.51 | 98.80 |
|  |  | BCI | 100.00 | 70.00 | 82.35 | 88.00 |
|  | OLI | ACMI | 100.00 | 91.00 | 95.29 | 96.40 |
|  |  | BCI | 100.00 | 77.00 | 87.01 | 90.80 |
| Xinjiang | ETM+ | ACMI | 100.00 | 90.00 | 94.74 | 96.00 |
|  |  | BCI | 96.81 | 91.00 | 93.81 | 95.20 |
|  | OLI | ACMI | 100.00 | 83.00 | 90.71 | 93.20 |
|  |  | BCI | 100 | 70.00 | 82.35 | 88.00 |
| Shanxi | TM | ACMI | 98.98 | 97.00 | 97.98 | 98.40 |
|  |  | BCI | 98.90 | 90.00 | 94.24 | 95.60 |
|  | OLI | ACMI | 95.88 | 93.00 | 94.42 | 95.60 |
|  |  | BCI | 98.51 | 66.00 | 79.04 | 86.00 |
| Inner Mongolia | TM | ACMI | 99.00 | 99.00 | 99.00 | 99.20 |
|  |  | BCI | 100.00 | 40.00 | 57.14 | 76.00 |
|  | OLI | ACMI | 100.00 | 87.00 | 93.05 | 94.80 |
|  |  | BCI | 100.00 | 67.00 | 80.24 | 86.80 |

## 5. Discussion

**5.1. ACMI performance in different Landsat sensors and coal types**

Landsat TM and ETM+ sensors sample the same regions of the spectrum, and they are often considered to be equivalent [24]. However, the differences in spectral response and some indices between OLI and prior Landsat sensors were reported to be subtle but notable [34]. Therefore, a TM/ETM+ image and an OLI image were selected for each study area (Table 2), to test the applicability of ACMI in the three Landsat sensors. According to the mapping accuracy summarized in Table 3, ACMI achieved the F1 score and OA greater than 0.90 for all the TM/ETM+ and OLI images. Further, generated through statistical averaging, the average F1 score of ACMI for TM/ETM+ and OLI are 0.97 and 0.95, respectively. In addition, the average OA of ACMI for TM/ETM+ and OLI are 97.40% and 96.00%, respectively. Overall, ACMI has achieved high accuracies for all the TM/ETM+ and OLI images, indicating the wide applicability of ACMI in different Landsat sensors.

The process of conversion of plant materials to coal is called coalification, producing different types of coal products such as lignite, subbituminous coal, bituminous coal, and anthracite [35]. Due to the variation in physical and chemical properties such as coalification degree, moisture and color, the spectral characteristics of EC of different coal types may be different. We therefore selected the six study areas that include various coal types (Table 1) to examine the robustness of ACMI in coal types. As shown in Table 3, BCI failed to identify almost all EC pixels in the study area of NRW, pointing to the bad performance of BCI in some regions of anthracite and bituminous coal. In contrast, ACMI showed its advantage of robustness to coal types with the average F1 score and OA no less than 0.93 and 94.60% for all coal types, respectively.

**5.2. Advantages and limitations of ACMI**

*5.2.1. Advantages of ACMI*

The ACMI is designed primarily to form a simple spectral enhancement approach that could enlarge the contrasts between EC surfaces and non-EC surfaces. Analysis of results suggests that ACMI performed much better than the commonly used methods. Overall, the ACMI has the following advantages: (1) It is highly automated and does not require any manual participation after data acquisition, making it much easier to transfer to other regions than classifier-based methods; (2) it has been proven to be robust to different landscape types, coal types and Landsat sensors; and (3) it has been proven to have a fairly stable optimal threshold and 0 can serve as its default threshold, which can eliminate the tedious steps of determining thresholds through sample training when conducting EC extraction for new regions. The above advantages make ACMI to be a reliable index for accurately and effectively extracting the EC distribution at a large scale.

*5.2.2. Limitations of ACMI*

Satellite remote sensing images are always affected by noise such as clouds and shadows, which may cause negative impacts on the classification or identification results [36]. To analyze the anti-noise ability of the ACMI, we used ACMI to extract EC from some cloud-contaminated Landsat images. From the results of these images, we found that some cloud shadows were misclassified as EC by the ACMI, such as those shown in Fig. 10. The Landsat OLI surface reflectance data in Fig. 10(a) was acquired on 24 June 2017 and its path/row in the WRS was 123/26. The misclassification of the cloud shadow may greatly accumulate in regard to time series or cloudy region applications. Therefore, it is necessary to select cloud-free images or to mask cloud shadows using the C Function of Mask (CFMask) algorithm [37] or the quality assessment (QA) band when applying the ACMI.

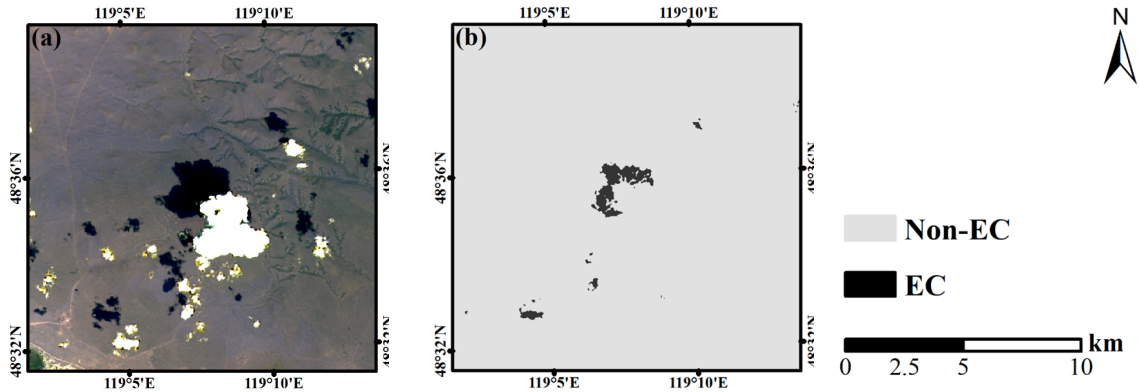

**Fig. 10.** An example of a cloud-contaminated OLI image. (a) is the true color synthesis image and (b) shows the EC extraction result of ACMI

Although the ACMI was examined under various background environments and coal types, a few factors that were not considered in this study may influence the accuracies of EC identification results. Seasonal and daily variations in the angle of the sun and changes in physical and chemical properties (eg., moisture and coalification degree) may affect the threshold and precision because of their possible influences on the reflectance patterns of EC surfaces. Therefore, the above factors may need to be considered when assessing the accuracies of EC identification methods.

Though ACMI achieved high accuracies for all study areas, two types of misclassifications occurred frequently. The first type is the sporadic or linear EC surfaces that were filtered as non-EC by the median filtering we used, even if they might be correctly classified by the ACMI formula (see (1)). The second type is the misclassifications that occur at the edges of EC polygons which are mainly due to (1) mixtures of EC and other land covers at edge pixels and (2) edge smoothing due to the use of median filtering.

In addition, the ACMI was only examined on Landsat images, its applicability may therefore need to be tested using images of other sensors. However, the ACMI cannot be used in some high-resolution satellite images like Chinese GF-1 because of their lack of SWIR1 and SWIR2 bands. Capturing high-resolution maps of EC is meaningful for reducing the influence of mixed pixels and thus enhancing the classification precisions. Therefore, it is necessary to develop a generic ACMI which can be applied in not only Landsat images but also high-resolution satellite images such as Chinese GF-1 and SPOT.

**5.3. The prospect of ACMI**

The free and open policy of some medium-resolution remote sensing resources, such as Landsat imagery, provides convenient opportunities to develop various methods to extract the spatiotemporal distribution of the specific land cover. The new index developed in this study contributes to the efforts being made to precisely identify EC surfaces based on Landsat imagery. This index uses a simple technique to summarize the common spectral characteristics of EC surfaces across the world and thus suppresses the background land covers to become a globally applicable index. Due to the wide applicability and high precision of the ACMI for identifying EC distributions, it can promote the application of relevant methods using EC surfaces as the medium, such as the methods proposed by Yang et al. [8] and Xiao et al. [9] for identifying damaged lands caused by coal mining, to national and even global scales. Further, given that 0 can serve as the default threshold of ACMI, it can also promote the automation degree of the above methods. In addition, EC surfaces are a reliable landmark of mining activities, making ACMI helpful for detecting illegal mining by identifying mining-induced EC distributions outside mining permits.

**6. Conclusions**

Coal is an important energy source and has contributed greatly to the development of human society. However, coal mining is criticized for leading to a lot of environmental issues. Mapping the EC distribution is significant in detecting coal mining activities and understanding environmental impacts. Therefore, this study aims to provide a method of extracting EC surfaces from multispectral satellite remote sensing imagery. Using Landsat TM/ETM+/OLI imagery, we proposed a new Automated Coal Mapping Index (ACMI) by increasing spectral separability between EC and non-EC surfaces. Six study areas, which contain different coal types and backgrounds, were selected to test the performance of ACMI. The EC mapping results showed that ACMI performed better than BCI in extracting EC pixels and suppressing non-EC surfaces. The accuracy assessment indicates that ACMI has shown better overall performance than BCI across all the study areas. The average UA, PA, F1 score and OA of ACMI are 99.24% 92.50%, 0.96 and 96.70%, respectively. In addition, ACMI was demonstrated to have a stable optimal threshold and 0 can serve as its default threshold of EC mapping. Since EC surfaces are directly related to coal mining activities, ACMI would be a reliable index for mapping the spatiotemporal dynamics of the footprints of coal mining activities.